# Discovering substantive disagreement with review articles?

**Rosati, Domenic**

**Simboli, Brian**

scite.ai, U.S.A | dom@scite.ai

Lehigh University, U.S.A

## INTRODUCTION

Disagreements help drive science. How does one identify and track them in scholarly literature? We ask the research question will searching review articles (RA) will be more time efficient for this purpose than searching non-review ones (NRA). This is especially so to the extent NRAs exceed RAs in a given field. We also discuss a metric for whether RAs report more substantive disagreements than NRAs.

This area of research is of interest given Blümel's et al. (2020) call for renewed interest in RAs. It may also be of interest to persons who want to identify scientific disagreements and track their resolution, whether researchers, journalists, students writing class papers, or specialists in the "science of science" and scientific communication. Attentiveness to tensions within a field should also be of interest to editors; as Moldwin et al. (2017) mention, "good review papers … help bring structure and understanding to the often disjointed and contradictory work that is at the forefront of a research field."

## METHOD

The hypothesis is that RAs discuss scholarly debates more so than NRAs. Data to evaluate the hypothesis uses 11,826,208 PubMed publication types grouped into RA including both reviews and systematic reviews (1,125,317) and NRA (10,700,891), abstracts and titles for those from Crossref, and bibliometric data about the references used by articles from scite.ai. These bibliometrics are whether references used by the article were supported or contrasted and the number of in-text citations to these references, number of citations to these references and the count of references from the article itself. In this work, the number of contrasting citations received by the references is considered a proxy for disagreement with the references and the number of in-text citations received is considered a proxy for engagement (cf. Hou, W.-R., 2011).

The first characterization method generates descriptive statistics for the bibliometrics of the references used by RA and NRA in the entire dataset. The second method develops a set of binary classification models for a subsample of the data to assess whether and to what degree those features predict review versus non-review articles. For classification, we sampled 100,000 publications from the initial dataset creating an 80,000 article training set and 20,000 article test set where the distribution of RA to NRA remained the same as the original.

To determine the predictive power of the bibliometric dimensions of the references selected by an article over other typical features used for classification, we trained a set of baseline models that only used the title and abstract and compared those against the use of bibliometrics features only or in conjunction with title and abstract. All of the models were linear classifiers built on features from: whether "review" was in the title or abstract; TFIDF on titles and abstracts; SPECTER embeddings (Cohan et al., 2020); or embeddings from DeBERTa v3 (He et al., 2021) fine-tuned on titles and abstracts from the dataset. Model training proceeded by testing bibliographic features alone and conjointly with these baseline models. Since language features in SPECTER and DeBERTa overpowered the bibliographic features we jointly trained an encoder from the language model with an encoder for bibliometric features, BiblioEncoder, that projected title and abstracts and bibliometric features into the same number of dimensions. In addition to the recall, precision, and F1 scores of each approach we inspected the coefficients in the trained models to give further insights into the predictive significance of each feature.

## RESULTS

As expected, the most noticeable difference between RA and NRAs are the reference counts with a median count of 73 references per RA and 31 references per NRA. Because of this difference the other reported scores are normalized by reference count. The average normalized number of contrasting citations to references (1.1 for both RA and NRA) and supporting citations to references (8.2 for RA and 8.4 for NRA) did not differ very much. However, there are noticeable differences in the number of in-text citations and citations to the references of review articles with a median of 192 in-text citations to references used by RA and a median 179 in-text citations per reference used by NRA as well as a median of 236 citations to references used by RA versus 230 to NRA. From this, RA's references are engaged with more heavily than NRA. However, one confounding fact we did not control for is the age of the reference selected. If the reference is older, it would have more time for citations to accrue and therefore the differences could be explained by RA using older references. Future work will address attempt to both think of more confounding factors such as disciplinary differences and review types as well as address them in more sophisticated analysis.

For prediction, classifying RA or NRA based on whether review is in the title or abstract is a poor approach, achieving an F1 score of 17.5, mostly due to a poor recall score of 9.9 with a precision of 75.6 indicating that this lexical approach misses most review articles. Using a linear classifier on top of the bibliometric features is not that much better with 28.9 F1 score, again mostly due to a poor recall score of 18.1 indicating that bibliometric features *by themselves* are not that powerful for distinguishing review articles. However, when combined, the model achieves a 53 F1 score (79.4 precision and 39.7 recall). Our best performing models were fine tuning SPECTER and DeBERTa trained on bibliometric and language features from titles and abstracts. DeBERTa achieved an F1 score of 74.5 trained on titles and abstracts alone and 83.3 when trained with BiblioEncoder. SPECTER embeddings without fine tuning only achieved a 65.1 F1 score with titles, abstracts, and bibliometric features. While SPECTER fine-tuned on abstracts and titles alone archives the best result of all models of an 89.6 F1 score, adding the bibliometric encoder to SPECTER does improve the baseline score from 65.1 F1 score to 78.9 F1 score. The score of SPECTER fine-tuned is likely the positive effect of in-domain pre-training that Gururangan et al. (2020) has observed which overpowers the effects of BiblioEncoder. These results indicate that bibliometric features, especially in conjunction with langauge features, tend to help with the task of RA classification. Table 1 summaries these results.

| Model | F1 | Precision | Recall |
| --- | --- | --- | --- |
| References | 28.9 | 72.2 | 18.1 |
| Title+Abstract | 17.5 | 75.6 | 9.9 |
| TA+Ref | 53.0 | 79.4 | 39.7 |
| TFIDF | 58.9 | **85.3** | 45.0 |
| TFIDF+Ref | 45.0 | 82.7 | 30.9 |
| SPECTER | 65.1 | 80.0 | 54.9 |
| SPECTER+Ref | 65.1 | 80.0 | 54.9 |
| SPECTER-Finetuned | **89.6** | **91.2** | **88.1** |
| SPECTER+BiblioEncoder | 78.9 | 82.1 | 75.9 |
| DeBERTa | 74.5 | 75.6 | 73.4 |
| DeBERTa+Ref | 74.4 | 74.9 | 73.9 |
| DeBERTa+BiblioEncoder | <u>83.3</u> | 84.7 | <u>82.0</u> |

**Table 1. Classifier performance on F1, Precision, and Recall scores for each model on the held out test set. Bold numbers are the best and underlined are the second best. Ref and BiblioEncoder indicate that bibliometric features of the references were used.**

When inspecting the features used by models trained to classify RA and NRA we find some notable predictors. Under the most interpretable model, using the review keyword in title and abstract as well as the references bibliometric features, the top 3 predictors of a review article by positive coefficient were in-text citations (1.81), average number of contrasting citations to each reference (1.23), and number of references used (0.85). For non-review articles, the top 3 predictors by negative coefficient were the total number of times a reference was neither supported nor contradicted (-1.76), average number of supporting citations per reference (-1.45), and average number of times a reference was neither supported nor contradicted (-1.02). This approach may suggest engagement with and disagreement to references (i.e., substantive disagreements) predict review articles while non-review articles might be best predicted by the uncontroversial references they use, or the number of supported references they used. Given the good performance of our classification models and relatively small differences in reference citation distributions, we speculate that future work should address granular characterization of review articles by review type and subject to identify if there are easier to identify and more prototypical RAs.

## DISCUSSION

These preliminary results indicate that it is more time efficient to search RAs than NRAs for disagreements. Still, questions arise requiring further work. First, we have assumed "review" has a stable, broadly applicable and generic meaning. Is this realistic? To the extent there are disciplinary differences in reliance on review articles (cf. Lamers et al. 2021), the answer may be "no". However, we must consider the multiplicity of standard RA types (Sutton et al. 2019). One reply is that the generic understanding of RAs suffices for very many purposes, since inclusion of filters for review articles in most scholarly databases is in response to this need. However, poor recall of the predictive models may indicate that only some types of review articles are easy to identify and prototypical. Another reply is that if practical needs arise, it is possible to apply the methods above by disciplinary fields and specific review types.

Second, do RAs typically cover more substantive disagreements than NRAs? Since RAs provide relatively high-level, synthetic literature reviews, we might expect this, at least for research conducted for some years. For practical purposes, though, it might be useful to create a metric that parses the meaning of "substantive" It would not replace subject specialist judgment. Instead, it would help in the construction of a practical information retrieval system that would rank for substantive disagreements. The metric would assign a score to RAs by weighing the following conjunction of their in-text citation characteristics: high engagement; number of citing items that contrast with each in-text citation: disagreement. The first measures the extent to which a proxy for "substantive" and the second measures disagreement. A retrieval mechanism can then sort RAs by the scores this metric assigns them. What functional form is appropriate for the metric is the subject for more research. Future work could understand if this was a useful system for retrieving high quality reviews articles.

A drawback is that focus on RAs excludes NRA reports of substantive disagreement, especially for very recent disagreements mentioned in NRAs but not yet in RAs. One can, however, mine for disagreements in recently published NRAs using the scoring metric just described. Also, one can use the metric to mine for disagreements reported in search sets containing both RAs and NRAs, though the sorted list of search results may be long, depending on the search topic. Future work will investigate these approaches beyond the biomedical domain and seek to provide a more comprehensive answer to whether review articles cover more substantive disagreements than non-review articles and better characterizations of specific types of reviews across disciplines. In addition, we intend to compare our developing work to that of Lamers et al. (2021) for the identification of disagreements themselves and we hope that our ability to quantify substantiveness adds additional nuance to the types of disagreements we can identify and where we can find them.